# Mode-matching metasurfaces: coherent reconstruction and multiplexing of surface waves

*Jiao Lin,*[*,†,] *Qian Wang,*[‡,§,∥] *Guanghui Yuan,*[∥,⊥] *Luping Du,*[∥] *Shan Shan Kou,*[†] *Xiao-Cong Yuan*[*,¶]

[†]School of Physics, University of Melbourne, VIC 3010, Australia

[‡]Optoelectronics Research Centre & Centre for Photonic Metamaterials, University of Southampton, Southampton, SO17 BJ, UK

[§]Institute of Materials Research and Engineering, Singapore 117602, Singapore

[∥]School of Electrical and Electronic Engineering, Nanyang Technological University, Singapore 639798, Singapore

[⊥]Centre for Disruptive Photonic Technologies, Nanyang Technological University, Singapore 637371, Singapore

[¶]Institute of Micro & Nano Optics, Shenzhen University, Shenzhen, 518060, China

**ABSTRACT:** Metasurfaces are promising two-dimensional metamaterials that are engineered to provide unique properties or functionalities absent in naturally occurring homogeneous surfaces. Here, we report a type of metasurface for tailored reconstruction of surface plasmon waves from light. The design is generic in a way that one can selectively generate different surface plasmon waves through simple variation of the wavelength or the polarization state of incident light. The ultra-thin metasurface demonstrated in this paper provides a versatile interface between the conventional free-space optics and a two-dimensional platform such as surface plasmonics.

Surface plasmon polaritons (SPPs) are coupled electromagnetic waves and charge density oscillations tightly confined at a metal-dielectric interface, which have created an appealing two-dimensional (2D) platform for designing high-performance optical components with small footprints in various applications,[1-3] and for bridging nanophotonics with semiconductor electronics.[4,5] Metallic surfaces patterned with subwavelength nanostructures, such as couplers[6-8] and planar optics (prism, lenses, etc),[9-12] have been developed for the generation and manipulation of SPPs. These engineered surfaces, also recognized as metasurfaces, are sometimes considered as the extension of bulk metamaterials into a 2D space, exhibiting novel functionalities and properties absent in naturally occurring homogeneous surfaces. Recently, new types of SPPs[13-15] have been demonstrated to possess interesting nondiffracting properties that can be promising for plasmonic manipulation of nanoparticles[16] and plasmonic circuits.[17]

Here, we report a generic design of metasurfaces that are capable of converting free-space light into an arbitrary type of SPP. Moreover, both wavelength and polarization multiplexing of SPPs can be realized by the metasurfaces in a straightforward manner, which enables selective excitation of a specific SPP mode by varying the wavelength or the polarization state of the incident light. The ultra-thin metasurface demonstrated in this paper provides a versatile interface between the conventional free-space optics and a two-dimensional platform such as surface plasmonics.

An SPP wave at a metal-dielectric interface is often simply described as a one-dimensional (1D) solution to the Helmholtz equation, neglecting the spatial variation in the transverse dimension. The out-of-plane electric field of such a simple monochromatic SPP wave propagating in the x-direction, for instance, can be expressed by $E_z(x,z) = A e^{-\alpha z} e^{i\beta x}$ (considering the dielectric half-space $z > 0$ ).[18] The parameter $\alpha$ indicates the confinement of the surface wave in the direction perpendicular to the surface, and the complex-valued propagation constant $\beta = k_0 \sqrt{\dfrac{\varepsilon_m \varepsilon_d}{\varepsilon_m + \varepsilon_d}}$, where $\varepsilon_m$, $\varepsilon_d$ are the permittivity of the metal and dielectric, respectively, and $k_0 = \omega/c$ is the wave vector in vacuum. $\alpha$ and $\beta$ are related through $\alpha^2 + \varepsilon_d k_0^2 = \beta^2$. Since the surface wave in general has a wavelength shorter than the free-space light of the same frequency, additional momentum is needed to fulfill momentum conservation for the excitation of the surface wave from free-space light. Conventionally, the condition can be met at a flat metal-dielectric interface with the help of attenuated total internal reflection from a denser optical medium (Supporting Information Fig. S1a).[19,20] Alternatively, the additional momentum can be provided by periodic corrugations created by a metasurface, such as 1D plasmonic gratings (extended infinitely in the other direction), which provides a more compact alternative (Supporting Information Fig. S1b).[21,22] In the latter case, the momentum matching condition

$$\mathrm{Re}\{\beta\} = k_0 \sin\theta + m\frac{2\pi}{\Lambda} \tag{1}$$

( $m$ is an integer number and $\theta$ is the incident angle) is achieved with the additional momentum parallel to the surface provided by the grating. By adjusting the period $\Lambda$ of the grating, the approach can accommodate a wide range of incident angles, including normal incidence, where the corresponding grating period equals to the wavelength of the SPP. Based on a 2D extension of the momentum-matching condition interpreted as mode-matching in this paper, we present a generic metasurface design consisting of two-dimensional (2D) plasmonic gratings (dubbed as mode-matching plasmonic gratings (MPGs)) for the coherent reconstruction of complex 2D SPP modes, and the transverse variation is no more neglected. In addition, we also show that the mode-matching metasurface design incorporates wavelength/polarization multiplexing of SPPs inherently.

Recently, the advent of the study of SPPs has revealed more intricate transverse profiles of the surface wave.[13-15, 23-27] Examples like plasmonic Airy beams (PABs),[13,14,23-25] have broadened our understanding of SPPs and opened up a whole new area because of the additional degree of freedom in the transverse dimension of SPPs. For its intriguing properties, such as propagating in a diffraction-free manner and following a parabolic trajectory,[13] the PAB is chosen as an exemplary surface wave to demonstrate the principle of our mode-matching metasurfaces. PABs were first experimentally produced by a periodic phase-shifted coupling grating milled into a metal surface, where the transverse field of the surface wave was reconstructed at a specific propagation distance.[14] Alternatively, light can be first coupled to a 1D SPP wave, and then through in-plane diffraction from a carefully designed aperiodic array, PABs can be created.[24] A controllable generation of PABs can be realized by projecting free-space Airy beams to a plasmonic grating with the help of a liquid-crystal spatial light modulator.[23] In this paper, we show that by matching the 2D field distribution of a known SPP mode e.g. a PAB in our case, a generalized plasmonic metasurface can be devised to convert a planewave into the desired 2D surface wave directly.

The 2D distribution (at a given height above the metal-dielectric interface) of the $E_z$ component of a PAB propagating in the *x*-direction is given by

$$E_z(u,v) = A_i(v - u^2 + i2au)\exp(av - 2au^2)\exp\left[i(uv + a^2u - \tfrac{2}{3}u^3)\right] \quad (2)$$

with the normalized coordinates defined as $u = \dfrac{x}{2\beta y_0^2}$ and $v = \dfrac{y}{y_0}$, where the parameter $a$ measures the apodization, $y_0$ is a transverse scaling factor, and $A_i(\text{g})$ is the Airy function.[13] Having the amplitude and phase distributions given by Eq. 2, in our design (Fig. 1a), a series of subwavelength slits are milled through an optically opaque Ag film at places where the phase of the desired surface wave (e.g. a PAB in the example) has the same value (e.g. zero) and the amplitude is above a threshold (Fig. 1b,c). When the perforated metal film is illuminated with a *coherent* source (e.g. a laser beam) with the correct polarization, each slit acts like a tiny antenna that receives the free-space light, and radiates surface wavelets (with the maximum radiation intensity occurring in the direction along the short axis of an slit). All antennas in the film are synchronized (in phase) because of the coherence of the incident planewave. However, in order for the surface wavelets emitted by all the antennas from different locations to interfere constructively forming the desired surface wave, these synchronized antennas must be placed at the wavefronts of the target surface wave. The aperture antennas (slits) are fabricated in an Ag film using focused ion beam (FIB) milling (Fig. 1d) and the reconstructed near-field intensity distributions (Figs. 1e,f) are found to be in a good agreement with the theoretically predicted profile of the desired PAB (Supporting Information Fig. S2).

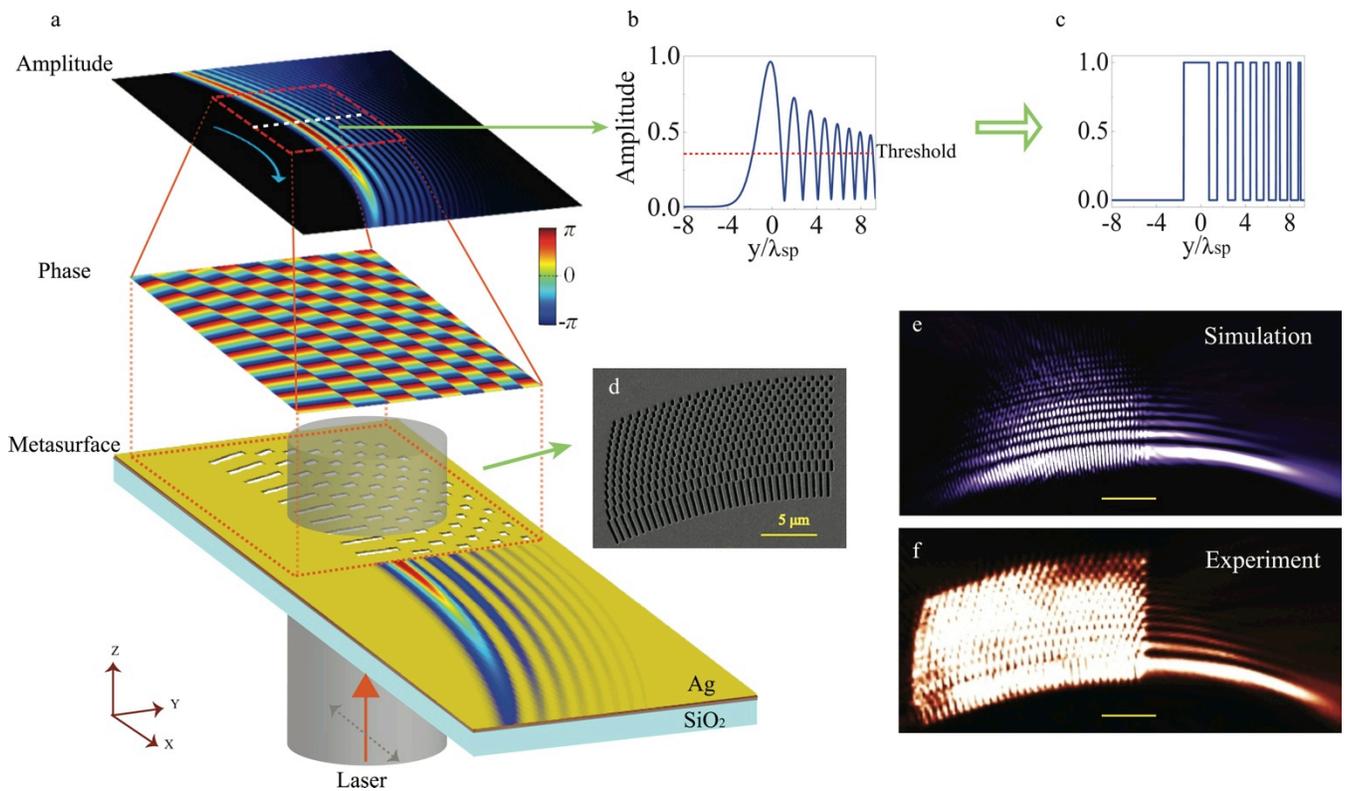

**Figure 1.** Mode-matching metasurface and the reconstruction of a surface wave. (a) Design of the MPG for the reconstruction of a PAB. The normalized amplitude of the 2D field distribution ($E_z$) of the target PAB is replaced with binary values "1" and "0". A threshold value is chosen for the quantization process as illustrated in (b) & (c). An optically thick Ag film (200nm) on a transparent glass substrate is perforated by FIB milling at positions where the phase of the 2D field distribution is equal to zero, and the binary amplitude is equal to unity. The resultant pattern consists of a series of slits with the width of ~200nm. When the metasurface is illuminated with a weakly focused *x*-polarized Gaussian laser beam (633nm and at matching frequency with the target SPP mode) from below, the SPP wavelets excited at individual slits propagate along the surface and add up constructively everywhere since they fulfill the local momentum matching condition for the specific surface wave. As a result, the target SPP wave is observed at the boundary of the perforated area. (b) The normalized amplitude profile of the PAB at a specific propagation distance. (c) The binary amplitude profile after thresholding. (d) Scanning electron micrograph of the metasurface that is used to reconstruct the PAB ($\lambda_{sp} = 613nm$, $a = 0.02$, and $y_0 = 600nm$). (e) 2D intensity distribution of the $E_z$ component of the reconstructed PAB obtained by full-wave calculation with the finite-difference time-domain (FDTD) method (scale bar: 5*μm*). (f) Near-field intensity distribution of the reconstructed PAB measured by near-field scanning optical microscopy (scale bar: 5*μm*).

Although a 1D periodic grating generates 1D SPPs (Supporting Information Fig. S1b), in a more general case like the reconstruction of PABs, the wavefront of the surface wave is usually not a straight line, which implies the space-variant orientation of the wavevector (gradient of phase change)—the vectorial nature of a momentum. Hence, *locally* the design of the grating must have not only the right periodicity (as indicated in Eq. 1) but also the correct orientation (as indicated by the direction of the local wave-vector of the target surface wave), such that the additional momentum provided by the grating fulfills the momentum-matching condition in both the magnitude and the direction simultaneously. Therefore, a simple replication of the structure in one dimension cannot guarantee the reconstruction of a surface wave generally. A direct comparison (Supporting Information Fig. S3) of the PAB reconstructed from an MPG with that from a 1D repetitive structure confirms the necessity of matching the plasmonic gratings with the *space-variant* wavefront of a propagating surface wave.

One interesting property of the mode-matching metasurfaces made of MPGs is the inherent multiplexing capability using incident angles or wavelengths. The angular multiplexing and wavelength multiplexing are usually interchangeable as shown in the momentum matching condition (Eq. 1). Because of the normal incidence configuration of the near-field scanning optical microscope (NSOM), we choose to demonstrate the wavelength multiplexing of SPPs. In Fig. 2, we show that the reconstructed SPP from a composite MPG, a metasurface consisting of several MPGs, varies with the incident wavelength. When the metasurface is illuminated by a coherent source of an individual wavelength, only the wavelets generated by all the slits of the corresponding constituent MPG can interfere constructively as the momentum matching condition is fulfilled by the particular incident wavelength. Therefore, the change of incident wavelength would result in the reconstruction of a different SPP mode. Given the parabolic trajectory of a PAB as shown in the example, the information carried by light of various wavelengths in free space will be directed to different direction on the metasurface.

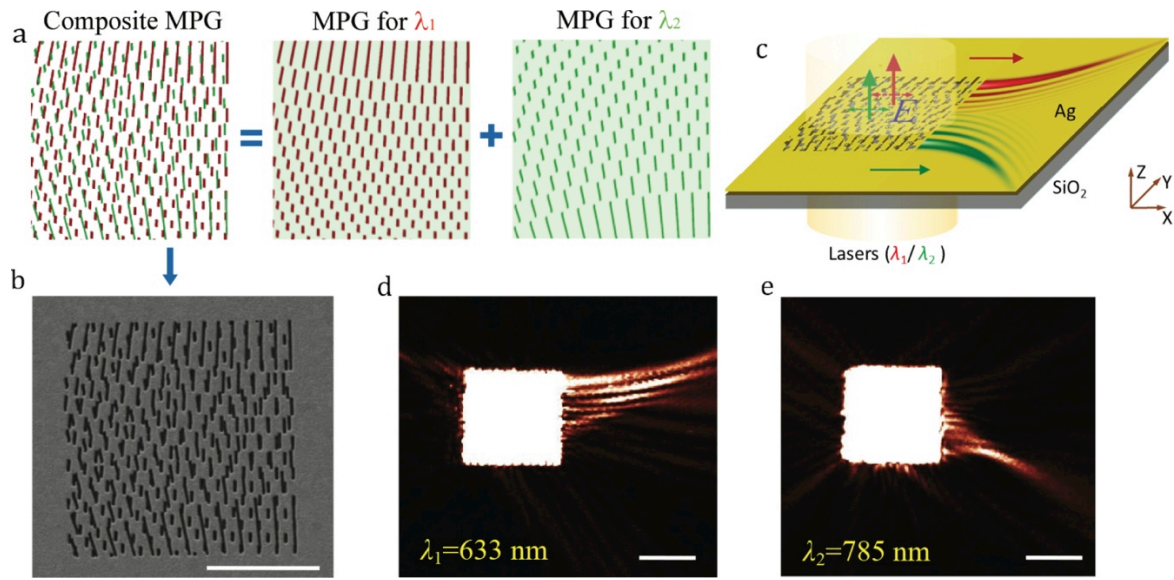

**Figure 2.** Wavelength-multiplexing with a mode-matching metasurface. (a) Design of the composite MPG for the reconstruction of two PABs of different effective wavelengths (613nm and 770nm at an Ag/air interface). The composite MPG is the superposition of two MPGs of different periodicity that are designed individually for reconstructing the two PABs. (b) Scanning electron micrograph of the metasurface for wavelength multiplexing (scale bar: 5$\mu m$). (c) Two PABs are reconstructed individually under the illumination of different laser sources (633nm and 785nm). (d) and (e), Near-field intensity distributions (scale bar: 5$\mu m$) of the two reconstructed PABs are measured by NSOM as in d ( $\lambda_{sp} = 613nm$, $a = 0.02$, and $y_0 = -600nm$) and e ( $\lambda_{sp} = 770nm$, $a = 0.02$, and $y_0 = 600nm$).

Another useful feature of the MPG-enabled metasurfaces is the capability of polarization multiplexing of SPPs. Since the constituent slits radiate SPPs at the highest efficiency when the polarization of the incident light is oriented along their short axes, the metasurface is sensitive to the polarization state of light. This intrinsic characteristic is explored in our experiment to selectively excite specific constituent MPGs in a composite MPG by varying the orientation of the linear polarization state of incident light (Fig. 3). In this case, a mode-matching metasurface is made by overlapping two MPGs sensitive to orthogonal polarization states. One of the constituent MPGs is excited when the incident linear polarization state is set to be perpendicular to its slits such that the corresponding surface wave will be reconstructed.

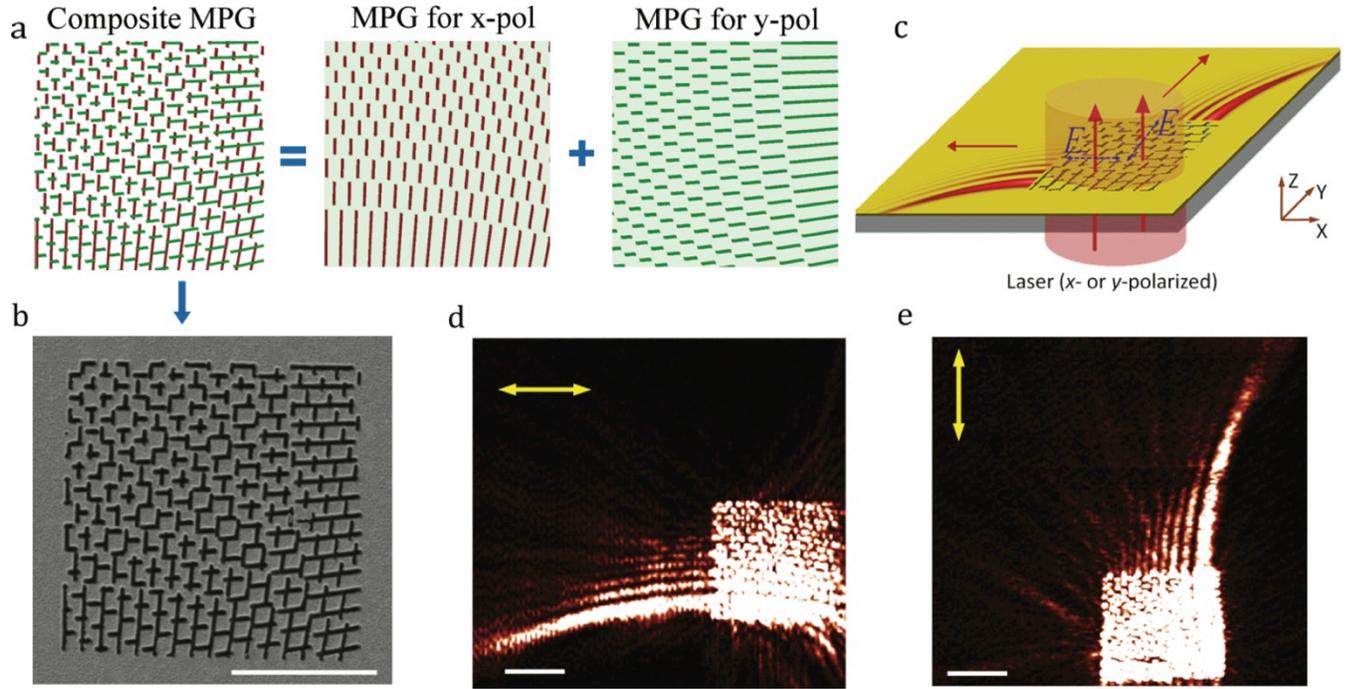

**Figure 3.** Polarization-multiplexing with a mode-matching metasurface. (a) Design of the composite MPG for the reconstruction of PABs at two incident orthogonal polarization states (*x*- and *y*-polarized). The composite MPG consists of two separate MPGs designed for orthogonal polarization states. (b) Scanning electron micrograph of the metasurface for polarization multiplexing (scale bar: 5$\mu m$). (c) The metasurface is illuminated with a linearly polarized laser beam of 633nm. The polarization state of the incident beam is changed by rotating a half-wave plate inserted in the optical path. (d) and (e), Near-field intensity distributions (scale bar: 5$\mu m$) of the reconstructed PABs as measured in d ($\lambda_{sp} = 613nm$, $a = 0.02$, and $y_0 = -600nm$) and in (e) ($\lambda_{sp} = 613nm$, $a = 0.02$, and $y_0 = 600nm$) with *x*-polarized and *y*-polarized incident light, respectively. The yellow arrows indicate the polarization state of the incident light.

The interference pattern (Supporting Information Fig. S4a) between a plane wave and a target surface wave (of the same frequency) resembles the design of single MPG (Supporting Information Fig. S4c) since both highlight the places with the same phase value. A surface hologram resulted from the interference can be understood as a 2D analogue to the volume holography. The planar interference pattern can be either recorded by a photoresist layer placed on a metallic film[28] or directly transferred into the metallic film by FIB milling.[29] Certainly, a surface hologram can also be used to reconstruct complex surface waves, e.g. the PABs (Supporting Information Fig. S4b). In contrast to the holographic technique, the MPG-enabled metasurfaces as proposed in the paper do not require any interference process, and can be seen as a generalized methodology to satisfy the space-variant momentum matching condition for the reconstruction of a surface wave. Potentially, a direct method like this can be extended to 3D space (e.g. metamaterials) by matching the field distribution in a volume but the realization of required subwavelength features in 3D must be accompanied by technological advancement of 3D nanofabrication, which is still under development. However, the readily available planar nanofabrication technologies such as photolithography, E-beam lithography and FIB ensure that the modular metasurface design concept can be exploited to its maximum capacity for harnessing surface waves.

In conclusion, we have demonstrated a straightforward method to reconstruct intricate surface waves based on generalized space-variant momentum matching. This enables the access to the additional degree of freedom in surface optics, e.g. the transverse dimension of SPPs. The mode-matching metasurfaces will also allow one to implement various multiplexing techniques to increase the capacity of plasmonic devices of a given size. In addition, the capability of selectively exciting a specific SPP transverse eigenmode by varying the parameters of the incident coherent radiation paves the way towards a more sophisticated interface for information routing from free-space optics to plasmonic circuits.

ASSOCIATED CONTENT

**Supporting Information**. General configurations for excitation of surface plasmon polaritons (SPPs); Transverse intensity profile of the reconstructed PAB as compared to the theoretical prediction; Comparison of the PAB reconstructed from an MPG with that from a repetitive structure; Comparison of the MPG with the holographic design. This material is available free of charge via the Internet at http://pubs.acs.org.

AUTHOR INFORMATION

**Corresponding Authors**

*Email: jiao.lin@osamember.org. *Email: xcyuan@szu.edu.cn

**Author Contributions**

J. Lin and Q. Wang contributed equally.

ACKNOWLEDGMENT

J. Lin and S. S. Kou are recipients of the Discover Early Career Researcher Award funded by the Australian Research Council under projects DE130100954 and DE120102352, respectively. J. Lin acknowledges the financial support from the Defence Science Institute, Australia. Q. Wang acknowledges the fellowship support from the Agency for Science, Technology and Research, Singapore. S. S. Kou acknowledges the financial support from the Melbourne Collaboration Grant and the Interdisciplinary Seed Fund through the Melbourne Materials Institute (MMI). X. Yuan acknowledges the support given by the National Natural Science Foundation of China under Grant Nos. 61036013 and 61138003, Ministry of Science and Technology of China under Grant No. 2009DFA52300 for China-Singapore collaborations, and National Research Foundation of Singapore under Grant No. NRF-G-CRP 2007-01. The authors would like to thank Dr Henry Cai (Carl Zeiss Pte Ltd, Singapore) for his assistance in preparing the samples.

# Supporting Information

Mode-matching metasurfaces: coherent reconstruction and multiplexing of surface waves

**Jiao Lin,**[†,⊥,*] **Qian Wang,**[‡,§,∥,‡] **Guanghui Yuan,**[∥,⊥] **Luping Du,**[∥] **Shan Shan Kou,**[†] **Xiao-Cong Yuan**[¶*]

[†]School of Physics, University of Melbourne, VIC 3010, Australia

[‡]Optoelectronics Research Centre & Centre for Photonic Metamaterials, University of Southampton, Southampton, SO17 BJ, UK

[§]Institute of Materials Research and Engineering, Singapore 117602, Singapore

[∥]School of Electrical and Electronic Engineering, Nanyang Technological University, Singapore 639798, Singapore

[⊥]Centre for Disruptive Photonic Technologies, Nanyang Technological University, Singapore 637371, Singapore

[¶]Institute of Micro & Nano Optics, Shenzhen University, Shenzhen, 518060, China

[‡]These authors contributed equally to this work

[*]Corresponding authors: jiao.lin@osamember.org or xcyuan@szu.edu.cn

# 1. General configurations for excitation of surface plasmon polaritons (SPPs)

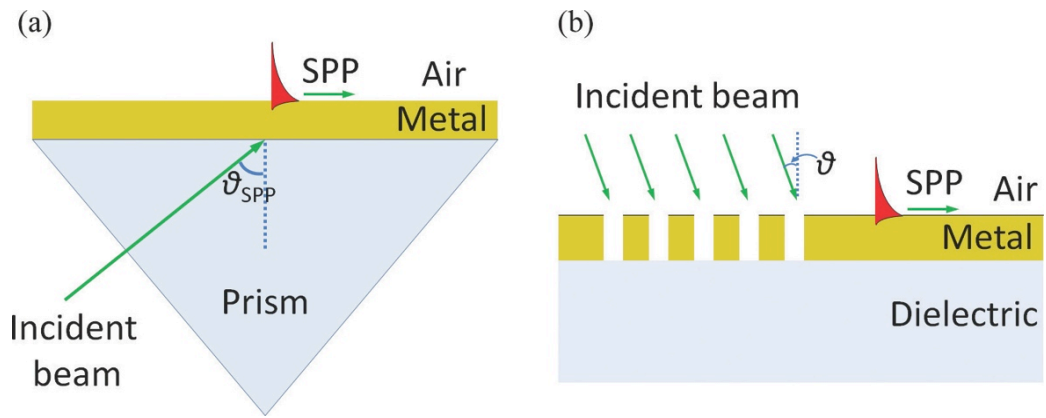

**Figure S1**. Configurations for the excitation of SPPs. (a) Excitation of SPPs at an air/metal interface with a prism. The thickness of the metal layer is typically at the range of tens nanometers in order to ensure the effective coupling of the electromagnetic field from one side to the other. (b) Excitation of SPPs with the help of a grating at the interface. The periodicity of the grating is designed to accommodate the incident angle of the light.

# 2. Transverse intensity profile of the reconstructed PAB as compared to the theoretical prediction

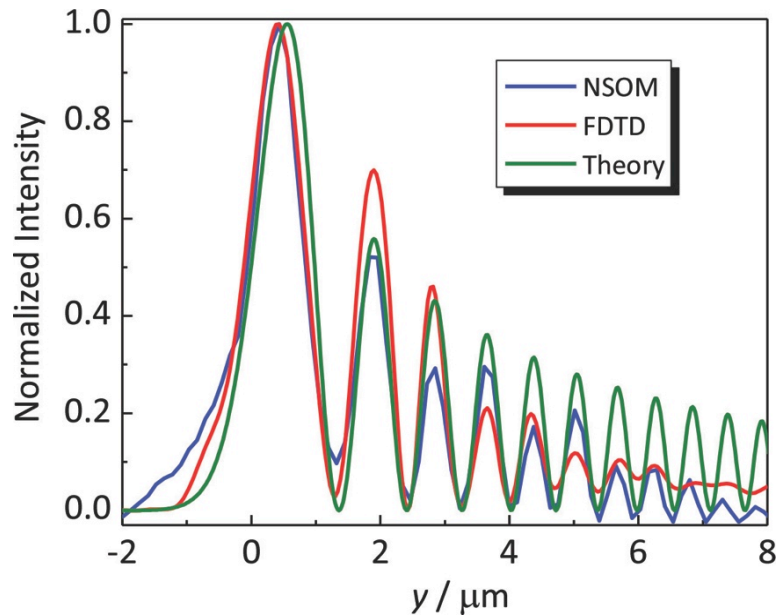

**Figure S2**. The transverse intensity profile of the reconstructed PAB as compared to the theoretical prediction. The curves are obtained at the propagation distance $x = 2\mu m$ by near-field scanning optical microscopy in the experiment (in blue), finite difference time domain (FDTD) calculation (in red) and the theoretical expression in Eq. 2 (in green). The parameters of the PAB are $\lambda_{sp} = 613 nm$, $a = 0.02$, and $y_0 = 600 nm$.

## 3. Comparison of the PAB reconstructed from an MPG with that from a repetitive structure

Ideally, a single structure ($n = 1$) is sufficient to reconstruct the PAB in both cases though the efficiency would be poor due to the small cross-section as compared to the incident laser beam. However, a simple repetition of the structure does not offer the constructive interference among the individual PABs they generated. A more noticeable deviation from the theoretical curve is observed when the number of periods increases from 5 to 10. For instance, although the PABs generated by the first and last periods of the grating are identical, there is always a traverse dislocation between them because of the well-known parabolic propagation trajectory of PABs. Therefore, the SPP generated by a longer grating deviates more from the target PAB because of the larger difference among propagation distances of the constituent wavelets. For example, in this case, the wavelet emitted by the last period travels longer to catch up with the one emitted by the first period, which means a larger lateral offset between this two wavelet.

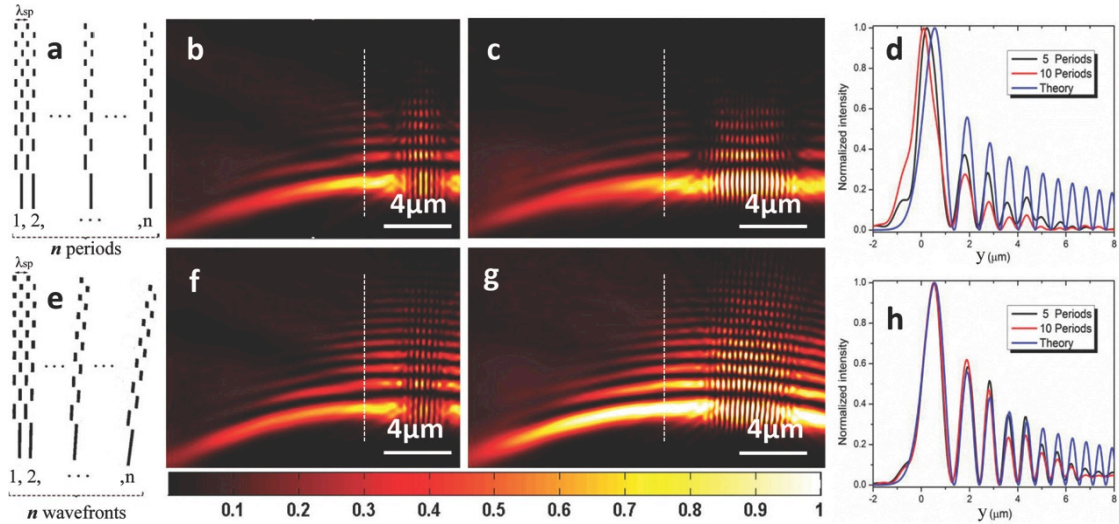

**Figure S3.** Reconstruction of PABs by the use of plasmonic gratings with various number of periods. (a) A grating formed by repeating a single structure (obtained from the wavefront at a specific propagation distance) $n$ times in the propagation direction. The pitch of the grating is equal to the wavelength $\lambda_{sp}$ of the SPP. (b)&(c), The near-field intensity distribution of the reconstructed surface wave from design (a) when $n = 5$ (b) and $n = 10$ (c), respectively. (d) The corresponding transverse intensity profiles at $x = 2\mu m$ for design (a). (e) Design of an MPG matched to the field of a PAB covering $n$ wavefronts. The structure is naturally curved as a result of the parabolic trajectory of the PAB. (f)&(g) The near-field intensity distribution of the reconstructed surface wave from design (e) when $n = 5$ (f) and $n = 10$ (g), respectively. (h) The corresponding transverse intensity profiles at $x = 2\mu m$ for design (e). The surface waves reconstructed from the MPGs are in better agreement with the target PAB than by using design (a). And the increase of the size of the MPG

(number of the wavefronts) shows a minimum effect on the fidelity of the reconstructed wave.

## 4. Comparison of the MPG with the holographic design

A good agreement between the SPP reconstructed from the MPG and the theoretical prediction of an ideal PAB (Eq. 2) is found whereas the SPP reconstructed from the hologram deviates from the theory. This can be attributed to the fact that the constituent aperture antennas (the parts in white) in the hologram vary in the width creating an uneven phase response across the device when illuminated with a planewave. On the contrary, the aperture antennas in the MPG have similar width and therefore free from the additional phase modulation giving a better reconstruction.

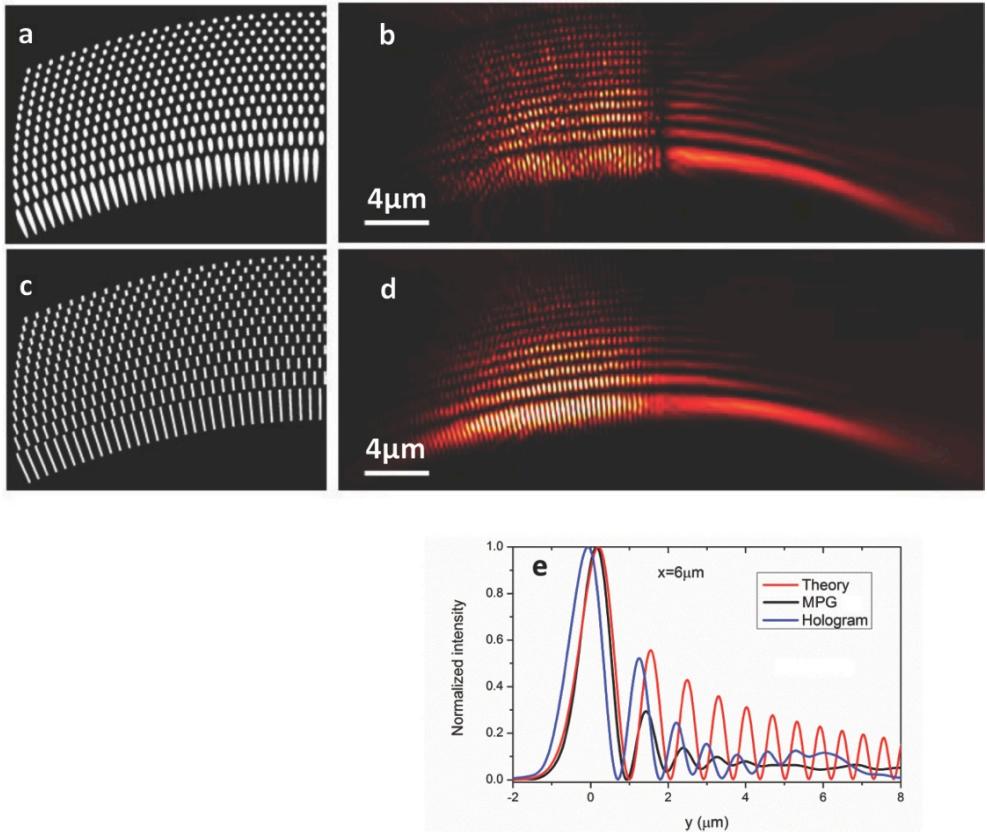

**Figure S4**. A surface hologram as compared with an MPG for the reconstruction of a PAB. (a) The holographic design is obtained by interfering the target PAB with a free-space planewave with the same frequency. The resultant 2D intensity distribution is then converted into a binary pattern as indicated in the figure. The parts in white indicating the places where the interference is constructive will be perforated through an Ag film. (b) The 2D intensity distribution ($E_z$) of the reconstructed SPP is observed in the full-wave calculation when the hologram is illuminated with an Gaussian beam (horizontally polarized) coming out of the plane. (c) An MPG designed for the reconstruction of the same PAB. (d) The reconstructed SPP obtained under the same condition as in (b). (e) The transverse intensity profiles (at the propagation distance $x = 6\mu m$) of the reconstructed SPPs (blue: hologram and black: MPG) and the target PAB (red).